\documentclass[conference,a4paper]{iaria} 
\pdfoutput=1 

\usepackage[babel=true,english=american]{csquotes}
\usepackage[USenglish]{babel}
\usepackage[alwaysadjust]{paralist}
\usepackage[
  babel=true, 
  expansion=alltext,
  protrusion=alltext-nott, 
  nopatch=eqnum, 
  final 
]{microtype}
\usepackage{upquote}
\usepackage[zerostyle=b,scaled=.9]{newtxtt}
\DisableLigatures{encoding = T1, family = tt* }
\usepackage[shortcuts]{extdash} 
\usepackage{relsize} 

\usepackage{fontawesome} 
\usepackage{lipsum}      
\usepackage{booktabs}
\usepackage{amsmath,amssymb,amsfonts} 
\usepackage[capitalise,nameinlink,noabbrev]{cleveref}
\usepackage{stfloats} 
\fnbelowfloat 
\usepackage[bb=boondox,bbscaled=.95,cal=boondoxo]{mathalpha}

\usepackage{array}

\usepackage{xurl} 

\vbadness=\maxdimen
\usepackage{algorithm}
\usepackage{algpseudocode}

\usepackage{makecell}

\usepackage[nolist]{acronym} 

\begin{acronym}
\acro{aws}[AWS]{Amazon Web Services}
\acro{iac}[IaC]{Infrastructure-as-Code}
\end{acronym}

\title{Towards Global Multi-Cloud Strategies: Insights into AWS and Alibaba Cloud Synergy}

\author{
  \IEEEauthorblockN{
    ~\\[-0.3ex]
    Martin G.\ Zizler\IEEEauthorrefmark{1},
    Malte Prieß\IEEEauthorrefmark{2}\;\orcidlink{0009-0004-4626-2513},
    Christoph P.\ Neumann\IEEEauthorrefmark{1}\;\orcidlink{0000-0002-5936-631X}
    \\[0.3ex]~
  }
  \IEEEauthorblockA{\IEEEauthorrefmark{1}%
    Department of Electrical Engineering, Media and Computer Science\\
    Ostbayerische Technische Hochschule Amberg-Weiden, Amberg, Germany\\
    e-mail: {\tt$\lbrace$m.zizler1\,|\,c.neumann$\rbrace$@oth-aw.de}
  }
  \\[-1.3ex]~
  \IEEEauthorblockA{\IEEEauthorrefmark{2}%
    Faculty of Computer Science and Electrical Engineering \\
    Kiel University of Applied Sciences, Germany \\
    e-mail: \tt malte.priess@haw-kiel.de}
}

\usepackage{ifpdf}
\ifpdf
\pdftrue
\fi

\makeatletter
\let\blx@rerun@biber\relax
\makeatother

\makeatletter
\def\ps@IEEEtitlepagestyle{
	\def\@oddfoot{\mycopyrightnotice}
	\def\@evenfoot{}
}
\def\mycopyrightnotice{
	{\footnotesize
		\begin{minipage}{0.8\textwidth}
			\centering
			Please cite as: \fullcite{selfref}.
		\end{minipage}
	}
}
\makeatother
\usepackage[shortcuts]{extdash} 

\DeclareBibliographyCategory{selfref}
\addtocategory{selfref}{selfref}

\begin{document}

\maketitle

\begin{abstract}
Multi-cloud strategies are increasingly adopted by modern enterprises to improve agility and resilience and to reduce vendor lock-in. Integrating workloads across providers, such as Amazon Web Services (AWS) and Alibaba Cloud, remains challenging due to interoperability and migration issues. This paper presents a comparative analysis of AWS and Alibaba Cloud, focusing on architectural, service, and policy differences affecting workload migration. Using both provider-native and open source Infrastructure-as-Code tools, we conduct an exploratory case study about the migration of Internet of Things (IoT) workloads. The results highlight key technical trade-offs and best practices for secure multi-cloud deployments, offering guidance for organizations pursuing AWS and Alibaba Cloud interoperability.
\end{abstract}


\section{Introduction}

According to Statista analysts, AWS is currently the world’s leading Cloud Service Provider (CSP), while Alibaba Cloud ranks fourth worldwide \cite{Richter2025}.
In contrast, within mainland China, Alibaba Cloud holds the top position, as reported by Canalys \cite{canalys2025}.
As enterprises expand internationally, region-specific regulations and preferences drive adoption of alternative providers \cite{Seth2024Compliance}.
We specifically selected these two providers because bridging the global market leader (AWS) with the dominant provider in mainland China (Alibaba Cloud) represents a highly relevant, real-world challenge for multinational enterprises that is currently underrepresented in the literature.
Multi-cloud strategies enhance agility, cost efficiency, resilience, and compliance \cite{Alonso2023,Chatzithanasis2021,Seth2024Compliance}, helping businesses mitigate vendor lock-in and address diverse operational needs \cite{Petcu2013}.
However, integrating multiple providers introduces challenges due to differences in architectures, APIs, and services, complicating interoperability and workload portability \cite{Ranjan2014,Munteanu2014}.
Although prior research addresses general multi-cloud concepts, practical guidance for migrating workloads specifically between AWS and Alibaba Cloud remains limited.
While technical hurdles, such as feature gaps in managed services, can often be solved via replatforming and custom workarounds, the overarching and arguably larger challenge lies in navigating strict, legally binding regulatory environments, including data residency and cross-border transfer restrictions.

This paper presents strategies for deploying and migrating workloads across AWS and Alibaba Cloud, focusing on technical and operational challenges. Section \ref{s:sota} outlines the state of the art; Section \ref{s:methods} describes the methodology; Section \ref{s:results} details the comparative analysis and deployments; Section \ref{s:discussion} discusses key findings; Section \ref{s:conclusionAndFutureWork} concludes with main contributions and future directions.

\section{State of the Art} \label{s:sota}

Prior comparative studies have predominantly focused on AWS, Azure, and Google Cloud \cite{Li2010, Saraswat2020, Rajendran2016}, providing quantitative and qualitative benchmarks but often excluding Alibaba Cloud. Zhang et al.\ \cite{Zhang2019} addressed this gap through a qualitative case study, identifying core vendor competencies and service delivery mechanisms unique to Alibaba Cloud. 

The quantitative and qualitative evaluation methodologies established in these previous studies represent past successes in multi-cloud benchmarking. Our work reuses these foundational approaches but extends them to practical, real-world migration scenarios and technical interoperability involving Alibaba Cloud, which remains underrepresented.

\section{Methods} \label{s:methods}

This section details the methodological framework used to analyze, design, and empirically validate a multi-cloud strategy across AWS and Alibaba Cloud. Our approach integrates structured comparative analysis with an exploratory case study, explicitly addressing gaps identified in prior studies and leveraging insights from recent empirical research.

\subsection{Research Design}

We employed a mixed-method comparative and exploratory approach, as advocated for cloud provider evaluations \cite{Li2010}. Our methodology combines a targeted literature review to identify technical, operational, and architectural challenges in multi-cloud migration with practical experimentation to ensure findings are empirically grounded.

\subsection{Comparative Framework}

Building on the foundational approaches discussed in Section \ref{s:sota}, our comparative framework evaluates real-world migration scenarios and technical interoperability, emphasizing four domains: Global Infrastructure, Core Service Portfolio, API Usage, and Infrastructure-as-Code (IaC) tooling.

\subsection{Strategy Development}

The comparative insights provided the basis for developing a multi-cloud architectural strategy. Guided by reference architectures in the literature \cite[pp.~72--76]{Mulder2024}, we evaluated managed Virtual Machine (VM), container, and serverless models. Reflecting recent empirical work, such as Rajendran et al.\ \cite{Rajendran2023}, which underscores the importance of use-case-driven benchmarking, we selected a representative IoT workload for our Proof of Concept (PoC). A serverless-first strategy was adopted, supplemented by VMs where feature parity was lacking. This means that our approach is more akin to a replatforming approach rather than a simple rehosting or \enquote{lift and shift} approach \cite{Hussain2025}. Replatforming typically requires a higher technical complexity, which means that it can surface deeper migration complexities, involving a higher amount of managed services. To systematically assess migration overhead and feature coverage, we implemented both provider-native—Amazon Web Services Cloud Development Kit (AWS CDK) and Resource Orchestration Service Cloud Development Kit (ROS CDK)—and provider-agnostic (CDK for Terraform) Infrastructure-as-Code (IaC) tools.

\subsection{Proof-of-Concept Development}

To test our strategy, we designed and deployed a reference IoT application on AWS using AWS CDK, then migrated and adapted it for Alibaba Cloud with ROS CDK. Parallel definitions using CDK for Terraform provided an agnostic baseline for comparison. The implementation process, informed by best practices in IaC-driven migration \cite{Kyadasu2025, Achar2021}, included:
\begin{itemize}
    \item Defining and mapping equivalent resources and deployment steps for each provider,
    \item Adapting configurations and documenting feature gaps,
    \item Recording manual interventions required for successful migration.
\end{itemize}

\subsection{Evaluation Methodology}

We evaluated each deployment approach using both quantitative and qualitative criteria:
\begin{itemize}
    \item Portability: Ease of migrating workload definitions and configurations
    \item Operational Transparency: Ongoing management and troubleshooting
    \item Maintenance Effort: Codebase maintenance
    \item Performance: Where measurable, indicative metrics were collected
    \item Security: Aligning IAM/RAM policies and Authentication
\end{itemize}
All findings were recorded systematically, with special attention to points of friction and required workarounds, as recommended by prior multi-cloud migration studies \cite{Zhao2022, Yussupov2019}.

\section{Results} \label{s:results}

This section presents the outcomes of the systematic comparative analysis between AWS and Alibaba Cloud in Section \ref{s:results:comparison} and the exploratory case study in Section \ref{s:results:poc}.
The comparative analysis is based on vendor documentation and migration guides, which may introduce bias.
To minimize overreliance on these secondary sources, we implemented the practical migration of a representative workload using IaC approaches.

\subsection{Comparative Analysis}
\label{s:results:comparison}

\subsubsection{Global Infrastructure}
AWS maintains global reach with 36 regions and 114 availability zones as of mid-2025, delivering strong coverage in North America and Europe \cite{Amazon2025Infra}. Alibaba Cloud operates 29 regions and 87 availability zones, with its core strength in Greater China \cite{Alibaba2025Infra}. Both providers offer specialized partitions like AWS GovCloud or AWS China \cite{Amazon2025China} to accommodate regulatory or sovereignty requirements.

It is important to note that Alibaba Cloud maintains two distinct infrastructures: AlibabaCloud.com, which serves international regions (e.g., Singapore, Frankfurt, Silicon Valley), and Aliyun.com (e.g., Shanghai, Beijing, Hangzhou), which serves regions within mainland China. In compliance with Chinese regulatory requirements, both Alibaba platforms operate in isolation. While international users can provision or manage resources in mainland China regions through AlibabaCloud.com, they are subject to a different regulatory framework (see also table \ref{tab:complianceconstraints}, regarding \enquote{Cross-border transfer restrictions}, \enquote{Provider restrictions}, and \enquote{Data residency}).

\subsubsection{Core Service Portfolio}

\begin{figure*}[htbp]
    \centering
    \includegraphics[width=0.9\textheight,angle=90,origin=c]{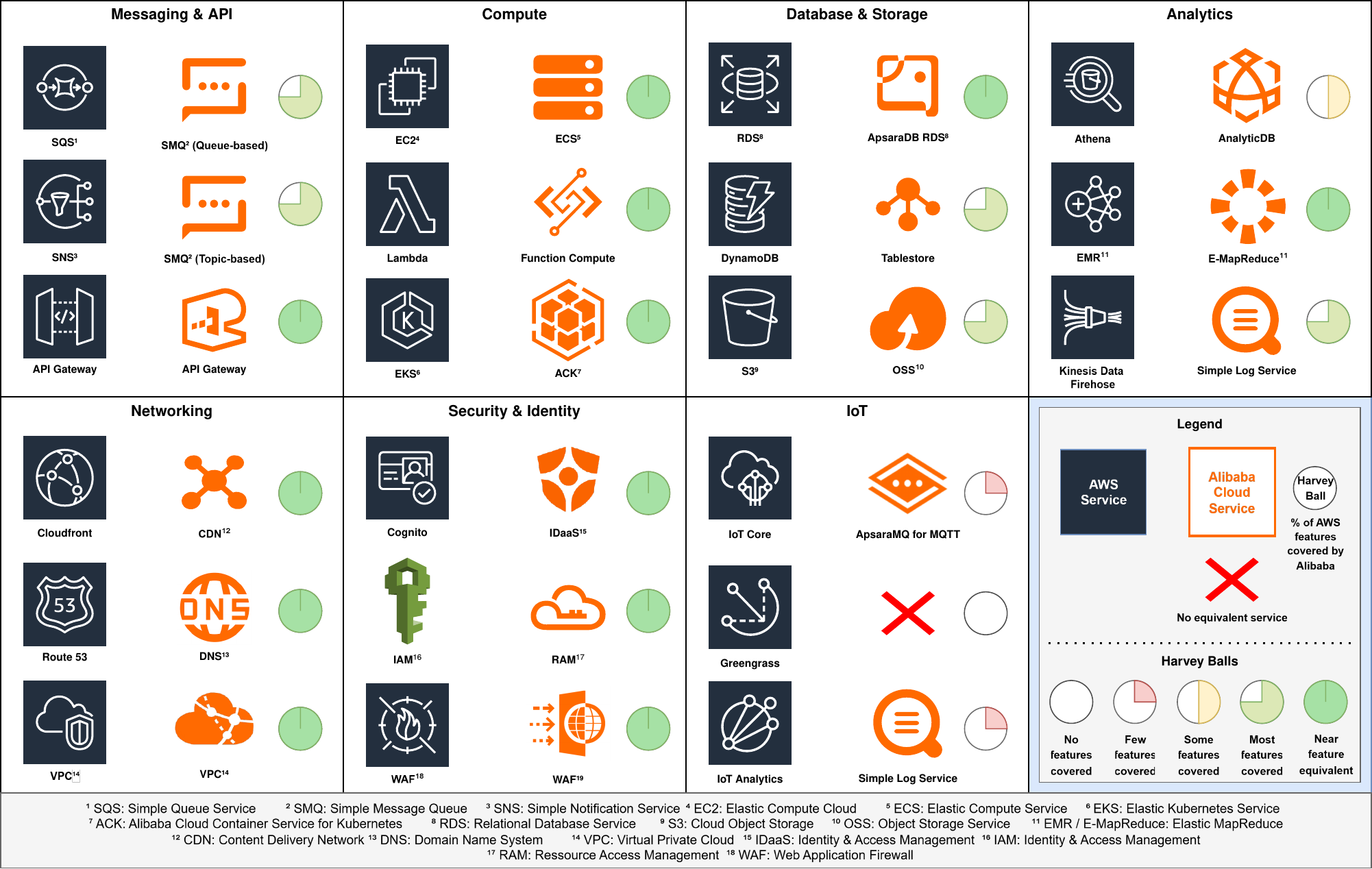}
    \caption{Cloud service overview and comparison between AWS and Alibaba Cloud core offerings, with a substantial gap in the IoT domain.}
    \label{fig:service-comparison}
\end{figure*}

Both AWS and Alibaba Cloud offer comparable core services across compute, storage, databases, networking, and security, though feature parity is not universal. Figure \ref{fig:service-comparison} provides an overview of the matched services and their functional completeness, based on vendor documentation \cite{Amazon2025, Alibaba2025} and a direct comparison by Alibaba \cite{AlibabaAWSServiceComparison}. Key differences are outlined below.

\subsubsection*{Messaging \& API Management}
Amazon Simple Queue Service (SQS) offers durable queuing with Standard (at-least-once) and FIFO (First-In-First-Out) modes, including dead-letter queues, short/long polling, and up to 14-day retention. Amazon Simple Notification Service (SNS) enables pub/sub delivery to SQS, Lambda, HTTP, Email, Mobile devices, SMS, Kinesis Data Firehose, and external providers (e.g., MongoDB). Alibaba Cloud Simple Message Queue (SMQ) supports queue-based and topic-based messaging with dead-letter queues, polling, and up to 7-day retention, but lacks FIFO mode. Its topic-based mode supports delivery to SMQ, Function Compute, HTTP, Email, SMS, and mobile endpoints. Both AWS API Gateway and Alibaba Cloud API Gateway are fully managed services that enable secure, scalable client-to-backend communication.

\subsubsection*{Compute}
AWS EC2 and Alibaba Cloud ECS provide flexible VM types. AWS Lambda and Alibaba Function Compute offer serverless, event-driven compute with auto-scaling and pay-per-use pricing. AWS EKS and Alibaba Cloud ACK deliver managed Kubernetes with high availability and reduced operational overhead.

\subsubsection*{Database \& Storage}
AWS Relational Database Service (RDS) and Alibaba Cloud ApsaraDB RDS can manage SQL databases. 
DynamoDB is a managed, serverless NoSQL database. Table Store delivers comparable features but has no true on-demand capacity mode, as it bills per Compute Unit (CU) instead of per request. 
AWS S3 and Alibaba Cloud Object Storage Service (OSS) are both fully managed object-storage services. S3 leads in the number of Storage classes. 

\subsubsection*{Analytics}
AWS Athena is serverless and lets you run SQL directly on S3 data for fast, ad-hoc analytics with no setup. Alibaba Cloud AnalyticDB provides batch processing and real-time analysis with support for both internal data and OSS, but requires cluster configuration. Athena is simpler to use, while AnalyticDB is more complex. 
AWS EMR and Alibaba E-MapReduce both run managed Hadoop and Spark clusters for big data processing in the cloud. 
AWS Kinesis Data Firehose and Alibaba Cloud Simple Log Service (SLS) both handle real-time data ingestion, transformation, and delivery to their cloud platforms. A key difference is that Firehose focuses on streaming data delivery, while SLS also includes built-in log analytics and monitoring features. 

\subsubsection*{Networking}
AWS CloudFront and Alibaba Cloud CDN both accelerate web content delivery via edge caching, reducing latency and improving performance. 
AWS Route 53 and Alibaba Cloud DNS provide scalable, globally distributed DNS management. 
Both clouds support secure, isolated virtual networks through their Virtual Private Cloud (VPC) services.

\subsubsection*{Security \& Identity}
AWS Cognito and Alibaba Cloud IDaaS both provide cloud-based user authentication and access management, integrating with their respective cloud services. 
AWS Identity and Access Management (IAM) and Alibaba Cloud Resource Access Management (RAM) offer the same core features of access management, including user, group, and role management, as well as permission controls. 
AWS Web Application Firewall (WAF) and Alibaba Cloud WAF both protect web apps from threats like SQL injection and XSS, offering customizable rules and real-time monitoring. 

\subsubsection*{IoT}
AWS IoT Core supports secure device connectivity, flexible protocols, and seamless integration with other AWS services. Alibaba Cloud ApsaraMQ for MQTT provides scalable MQTT messaging but lacks advanced device management and integration features found in IoT Core. 
AWS Greengrass offers edge computing for IoT, enabling local compute and sync when offline. Alibaba Cloud has no direct equivalent service to Greengrass. 
AWS IoT Analytics delivers managed pipelines for processing IoT data, while Alibaba Cloud lacks a truly equivalent service. Simple Log Service (SLS) can be used for basic data ingestion and analytics. 

\subsubsection{API Usage}
Both clouds expose RESTful APIs and SDKs covering major languages, but slightly differ in endpoint conventions and authentication depending on configuration. The API documentations for both CSPs show that basic API requests are still very similar across both (e.g., Bucket API documentation for Amazon S3 \cite{Amazon2025S3API} vs.\ Alibaba OSS \cite{Alibaba2025OSSAPI}).

\subsubsection{Infrastructure-as-Code Tools}
AWS CDK (CloudFormation) and Alibaba ROS CDK provide native IaC tooling. Meanwhile, Terraform or OpenTofu, as well as CDK for Terraform, which are popular for multi-cloud deployments, also support both CSPs \cite{Kyadasu2025}. Native IaC tooling generally provides faster support for new resource types and higher levels of abstraction. 

\subsection{Exploratory Case Study}
\label{s:results:poc}

\subsubsection{Workload and Architecture}
\begin{figure*}[htb!]
    \centering
    \includegraphics[width=0.9\linewidth]{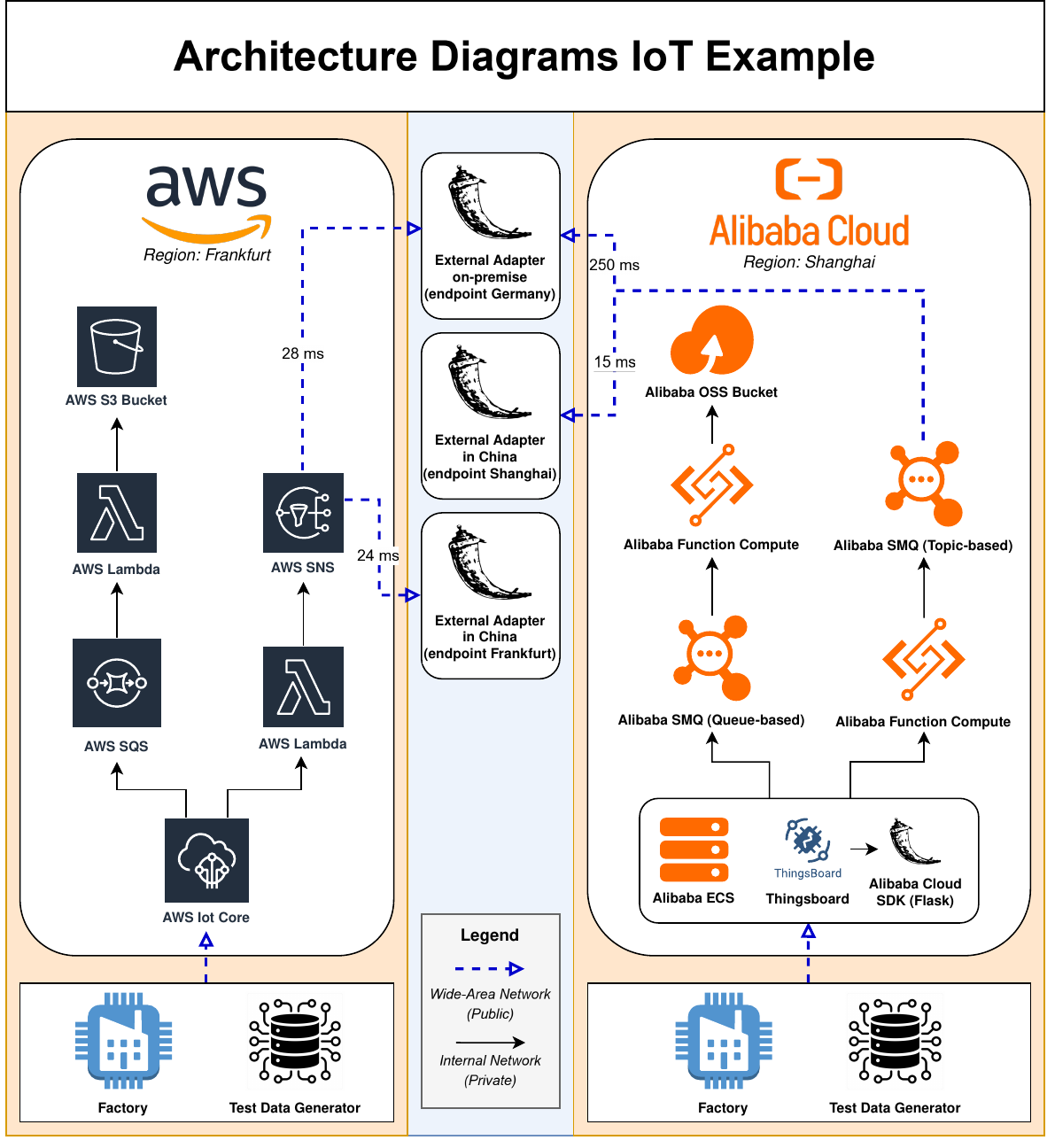}
    \caption{Architecture diagram showing resource mapping and data flow of a basic IoT Stack.}
    \label{fig:architecture-diagram}
\end{figure*}

As a Proof-of-Concept (PoC), a representative IoT workload consisting of compute, storage, and event-driven processing was implemented using both provider-native and agnostic IaC tools for deployment. The workload includes serverless functions, object storage buckets, event triggers, and messaging services, which can be seen in Figure~\ref{fig:architecture-diagram}. Equivalent resources were used for AWS and Alibaba Cloud. Additional adaptation was required for the AWS IoT Core. While ApsaraMQ for MQTT exists as a potential replacement, it is just a generic MQTT broker with a very sparse feature set (see \ref{s:results:comparison}). Therefore, Thingsboard was selected as an open alternative and deployed on Alibaba Cloud ECS. Similar to IoT Core, Thingsboard fully supports X.509 Certificate-based mutual authentication, which can be managed by device \cite{ThingsboardDocs}. It also supports custom Rule Chains to process events. To securely send data from a Thingsboard Rule Chain to other Alibaba Cloud services, a simple Flask server that can get access by utilizing the Alibaba Cloud SDK for Python was also added as an intermediary. Furthermore, an external adapter was set up to show that data can also be retrieved from the CSPs. This adapter was also used to test latency differences depending on deployment location.

\subsubsection{IaC Implementation}
AWS CDK (Python) and Alibaba ROS CDK (Python) were used to define and deploy the stack natively. Most of the resource definitions translated with little adaptation needed, as they have very high overlap, as shown in \ref{s:results:comparison}. The remaining required manual adaptation is due to differences in parameterization, IAM/RAM policy syntax, or missing features (e.g., managed IoT services).

CDK for Terraform (Python) was used to define stacks utilizing the same Python environment, targeting both AWS and Alibaba Cloud providers, while OpenTofu was used as the underlying IaC tool (open-source Terraform fork). Very similar to the provider-native approach, the Alibaba Cloud code base required provider-specific adaptation in all resources to accommodate differences in event sources and IAM/RAM, as well as configuration.

\subsubsection{Deployment and Operational Metrics}

For these metrics, the Alibaba Cloud equivalent of AWS IoT Core (an ECS-based Thingsboard deployment) is excluded from the IaC line count, as automating this setup would require substantial custom scripting. For fair comparison, AWS IoT Core is also omitted.

Table~\ref{tab:deploymenttimesandlinecounts} summarizes deployment and teardown times, as well as Lines of Code (LoC) required for IaC definitions of each approach. Deployment times were similar between provider-native and provider-agnostic tools within each platform, but Alibaba Cloud was faster. Line counts were measured using VS Code Counter with code formatted via Black to ensure consistency. Notably, the AWS CDK for Terraform (CDKTF) implementation required substantially more code than alternative approaches, primarily due to AWS’s detailed IAM model and the need for explicit resource linking. In contrast, the AWS CDK benefits from high-level constructs, resulting in a more concise codebase. Differences in code length among Alibaba Cloud tools were comparatively minor.

\begin{table}[ht]
\centering
\footnotesize
\begin{tabular}{llccc}
\toprule
\textbf{Platform} & \textbf{Tool} & \textbf{Deploy} & \textbf{Destroy} & \textbf{Lines} \\
\midrule
AWS           & CloudFormation & 1m 15s & 56s & 75 \\
AWS           & CDKTF          & 1m 11s & 30s & 184 \\
Alibaba Cloud & ROS            & 24s    & 26s & 146 \\
Alibaba Cloud & CDKTF          & 19s    & 26s & 143 \\
\bottomrule
\end{tabular}
\caption{Deployment Times, Teardown Times, and IaC Line Counts of IoT Stack.}
\label{tab:deploymenttimesandlinecounts}
\end{table}

The architecture diagram in Figure \ref{fig:architecture-diagram} also provides some latency numbers, annotated to the blue dashed arrows for wide-area networks.
Network transfers within the same country (Germany to Germany or China to China) have low latency.
Cross-border egress (from Alibaba in Shanghai to the external adapter in Germany) introduces a latency about ten times higher.
We provide an extended set of quantitative benchmarking in a publicly available Master's thesis \cite{zizler_martin_2025mt}.

\section{Evaluation}
\label{s:discussion}

This section summarizes the main findings from our comparative analysis and exploratory case study, highlighting key trade-offs of native versus agnostic IaC approaches.

\subsection{Interpreting the Comparative Analysis}
AWS and Alibaba Cloud both offer mature core services, but differ in regional coverage, service completeness, and compliance. IoT-heavy workloads on Alibaba Cloud require additional custom or third-party solutions to address service gaps, whereas AWS provides more integrated support.

\subsection{Evaluating the PoC}
The deployments allow evaluation of the following criteria:

\begin{enumerate}
  \item \textbf{Portability}: Provider-native IaC (AWS CDK, ROS CDK) offers rapid access to new features and high-level constructs, but poor cross-provider code reuse. Agnostic IaC (OpenTofu) enables a unified code base, but still needs extensive provider-specific adjustments.
  \item \textbf{Operational Transparency}: Native tools integrate better with CSP management interfaces, offering richer diagnostics and control, unlike OpenTofu-based stacks.
  \item \textbf{Maintenance Effort}: Unified CDK for Terraform code bases can reduce duplication but increase maintenance for provider plugin updates. Native stacks benefit from vendor-managed updates but require managing separate pipelines.
  \item \textbf{Performance}: Deployment times for Alibaba Cloud seem to be a bit faster (see Table \ref{tab:deploymenttimesandlinecounts}).
  \item \textbf{Security}: Ensuring least-privilege access required manual effort to align IAM (AWS) and RAM (Alibaba Cloud) policies. For instance, AWS CDK provides high-level abstractions for granting permissions (e.g., allowing Lambda to write to S3), whereas ROS CDK often necessitates explicit role and policy configuration. Additionally, synchronizing certificate-based authentication across providers involved adapting identity management to maintain secure communication over wide-area networks.
\end{enumerate}

\setlength{\extrarowheight}{4pt} 
\begin{table*}[htbp!]
\centering
\footnotesize 
\begin{tabular}{>{\centering\arraybackslash}p{2.2cm}p{3.8cm}>{\centering\arraybackslash}p{4.5cm}p{5.9cm}}
\toprule
\textbf{Constraint type} &  & \textbf{Source} & \textbf{Significance for multi-cloud} \\
\midrule
Data residency &
Transfer of personal data to third countries (outside the EU/EEA) is only permitted under specific conditions. &
{GDPR Art.~44--49 \cite{EU_GDPR}} &
Storing EU personal data in Alibaba Cloud Mainland may breach GDPR; transfers require adequacy decisions, SCCs, or legal exceptions. \\
Cross-border\par transfer restrictions &
Data exports from China may require prior security assessment and government approval. &
{DSL Art.~31--37 \cite{China_DSL}, \par
CSL Art.~37 \cite{China_CSL}, \par
MLPS~2.0 \cite{China_MLPS_2_0}} &
Transfers from Alibaba Cloud Mainland to AWS Frankfurt may require CAC approval and data export security review. \\
Tenant isolation &
Data belonging to different customers, departments, or patients must be kept logically and technically separated. &
{GoBD 2020 \cite{BMF_GoBD_2020}, \par
SOX §404 \cite{US_SOX}, \par
HIPAA 164.308(a)(4) \cite{US_HIPAA_164_308_a_4}} &
IaC should enforce resource separation (e.g., VPCs, IAM roles, storage buckets) per tenant to prevent data leakage. \\
Auditability &
Access to systems must be traceable and securely logged for compliance and incident analysis. &
{BSI C5:2020 (e.g., OPS-07) \cite{BSI_C5_2020}, \par
ISO 27001 8.15 (Logging) \cite{ISO_IEC_27001_2022}, \par
GDPR Art.~30,~33 \cite{EU_GDPR}} &
Cloud-native logging (e.g., AWS CloudTrail, Alibaba ActionTrail) should be enabled, retained, and protected. \\
Classification requirements &
Operators of critical infrastructure must classify systems and apply tiered protection accordingly. &
{NIS\,2 (EU 2022/2555, Art.~21) \cite{EU_NIS2}, \par
BSIG §8a \cite{BSIG} \& IT Security Act \cite{DE_IT_Sicherheitsgesetz}, \par
MLPS 2.0 \cite{China_MLPS_2_0}} &
Selection of certified services only (e.g., BSI C5) and onshore deployment; additional monitoring and emergency mechanisms if necessary; classification as a necessary prerequisite for protective measures. \\
Provider restrictions &
A cloud provider may be subject to foreign government access demands (e.g., US CLOUD Act, CN national laws). &
{CLOUD Act \cite{US_CLOUD_Act}, 
GDPR Art.~48 \cite{EU_GDPR}, \par
BSIG §9b \cite{BSIG} \& IT Security Act \cite{DE_IT_Sicherheitsgesetz}, \par
Gaia-X standards, if applicable \cite{GaiaX_PolicyRules_22_04, GaiaX_TrustFramework_22_10}} &
Onshore providers preferred to avoid extraterritorial access; regulatory context may disqualify US/CN providers for critical workloads; clarification on who can enforce access to data is essential. \\
Access and \par identity control &
Only authorized users should access data, using strong authentication and role-based access control. &
{GDPR Art.~32(1)(b) \cite{EU_GDPR}, 
ISO 27001 8 (Technological controls) \cite{ISO_IEC_27001_2022}, 
BSI C5 (e.g., IDM-09) \cite{BSI_C5_2020}, CSL Art.~21 \cite{China_CSL}} &
IAM (AWS) and RAM (Alibaba) should enforce RBAC, MFA, and auditable access policies. \\
Data minimiza-\par{}tion \& purpose limitation &
Only necessary data may be processed and stored for clearly defined purposes. &
{GDPR Art.~5(1)(c) \cite{EU_GDPR}, \par
PIPL Art.~6 \cite{China_PIPL}} &
IaC and pipelines should be limited to minimal datasets and clearly scoped processing goals. \\
\bottomrule
\end{tabular}
\caption{Examples of compliance constraints derived from international legal sources and industry-specific standards.}
\label{tab:complianceconstraints}
\end{table*}


\subsection{Compliance Considerations}

Compliance challenges are increased by regional regulations. For example, Alibaba Cloud’s mainland China partition is subject to local laws like the China Cybersecurity Law, requiring data residency and stricter controls on cross-border flows \cite{ChinaCrossborderInformation}. AWS China and Alibaba’s specialized regions address sovereignty but require careful architectural planning.

To further classify the regulatory requirements, table \ref{tab:complianceconstraints} shows examples of compliance constraints derived from international legal sources and industry-specific standards. The analysis does not aim to provide a concluding legal evaluation of regulatory frameworks; rather, it offers a first technical abstraction of selected requirements. From an architectural perspective, it illustrates how regulatory requirements, such as the GDPR, the German IT Security Act 2.0, PIPL, or the CLOUD Act, may translate into concrete technical design decisions, including role-based access control, client separation, and data localization. 

Due to the limited harmonization of international regulations, globally uniform cross-country infrastructures remain challenging. A more realistic horizon lies in compliance-aware, modular architectures that enable controlled interoperability while respecting regional legal constraints.

\subsection{Lessons Learned \& Best Practices}
\begin{itemize}
  \item Utilize a service mapping matrix to track equivalences.
  \item Use provider-native IaC services to make use of high-level abstraction and have a higher operational transparency. Use cloud-agnostic IaC to achieve more equal code bases between different CSPs.
  \item Plan for manual adaptation where services don't match.
  \item Leverage native security tools and audit access policies.
\end{itemize}

\subsection{Threats to Validity}
Our PoC focused on a basic IoT stack; results may not generalize to large-scale data processing, or CSP-specific managed services outside the evaluated scope. Pricing and performance data are indicative; real-world figures will vary by workload size, region, and time. Finally, CSP feature sets evolve rapidly, so this mid-2025 snapshot may differ from future states.

\section{Conclusion and Future Work} \label{s:conclusionAndFutureWork}

Our analysis compared AWS and Alibaba Cloud across infrastructure, services, Infrastructure-as-Code tools, and regulatory frameworks, with the findings validated through a small-scale IoT proof of concept.
AWS has a more mature service portfolio and leads in innovation speed. This can make it harder to develop a true multi-cloud strategy based on using serverless services.
Our comparative analysis extends previous studies to real-world migration scenarios and technical interoperability.
In conclusion, the paper closes several gaps in multi-cloud literature for global approaches that comprise China and Alibaba Cloud.

Several opportunities for further investigation present themselves moving forward:
\begin{itemize}
  \item Broader workloads with other managed services.
  \item Performance and cost benchmarking at scale.
  \item Explore integration of multi-cloud management platforms.
  \item Assess interoperability with third-party SaaS offerings.
  \item Evaluate specific privacy and security implications of using CSPs governed by distinct national legal frameworks (e.g., data sovereignty and state access concerns regarding Alibaba Cloud).
\end{itemize}

By advancing these areas, future work can further reduce operational friction and enhance the robustness of global multi-cloud deployments.

\begingroup
\sloppy
\printbibliography[notcategory=selfref]

@INPROCEEDINGS{selfref,
		title={{Towards Global Multi-Cloud Strategies: Insights into AWS and Alibaba Cloud Synergy}},
		author={Martin G. Zizler and Malte Prieß and Christoph P. Neumann},
		booktitle={Proc of the 17th International Conference on Cloud Computing, GRIDs, and Virtualization (Cloud Computing 2026)},
		_organization={International Academy, Research, and Industry Association (IARIA)},
		address = {Lisbon, Portugal},
		month=apr,
		year={2024},
		url = {https://www.thinkmind.org/library/CLOUD_COMPUTING/CLOUD_COMPUTING_2026/cloud_computing_2026_3_10_20015.html},
		issn = {2308\-/4294},
		pages = {34--41}
	}
\endgroup 

\end{document}